\documentclass[preprint,showpacs,preprintnumbers,amsmath,amssymb]{revtex4-1}
\usepackage{graphicx}
\usepackage{dcolumn}
\usepackage{bm}
\usepackage[latin1]{inputenc}

\usepackage[version=3]{mhchem}
\usepackage{amsfonts}
\usepackage{amssymb}
\usepackage{amsmath}

\begin{document}
\title{sp magnetism in clusters of gold-thiolates}
\author{A. Ayuela}
\email{swxayfea@sw.ehu.es}
\affiliation{
Centro de F\'{\i}sica de Materiales CFM-MPC CSIC-UPV/EHU, 
Donostia International Physics Center (DIPC),
 Departamento de F\'\i sica de Materiales,
Fac. de Qu\'\i micas, Univ. del Pais Vasco UPV-EHU, 20018
San Sebasti\'an, Spain}

\author{P. Crespo}
\email{pcrespo@adif.es}
\affiliation{Instituto de Magnetismo Aplicado, UCM-CSIC-ADIF. Las Rozas. P. O. Box 155, Madrid 28230 and Dpto. F\'\i sica de Materiales, Universidad Complutense}

\author{M.A. Garc\'\i a}
\email{magarcia@icv.csic.es}
\affiliation{Instituto de Cer\'amica y Vidrio. CSIC c/Kelsen,5  Madrid 28049}

\author{A. Hernando}
\email{antonio.hernando@adif.es}
\affiliation{Instituto de Magnetismo Aplicado, UCM-CSIC-ADIF. Las Rozas. P. O. Box 155, Madrid 28230 and Dpto. F\'\i sica de Materiales, Universidad Complutense}

\author{P. M. Echenique}
\email{wapetlap@sq.ehu.es}
\affiliation{
Centro de F\'{\i}sica de Materiales CFM-MPC CSIC-UPV/EHU, 
Donostia International Physics Center (DIPC),
 Departamento de F\'\i sica de Materiales,
Fac. de Qu{\'i}micas, Univ. del Pais Vasco UPV-EHU, 20018
San Sebasti\'an, Spain}



\begin{abstract}
Using calculations from first  principles, we herein consider the bond made between thiolate  with  a range of different Au clusters, with  a particular focus on the spin moments involved in each case.   For odd number of gold atoms, the clusters  show a spin moment of 1.~$\mu_B$.  The variation of spin moment with particle  size is particularly dramatic, with the spin moment being  zero for even numbers  of gold atoms.  This variation may be linked with changes  in the odd-even oscillations  that  occur  with the  number  of  gold  atoms,  and  is associated  with the formation  of a S-Au  bond. This  bond leads  to the presence of an extra electron that  is mainly sp in character in the  gold part.  Our results  suggest that  any thiolate-induced  magnetism that  occurs in gold nanoparticles  may be localized in a shell below the surface,  and can be  controlled by modifying the coverage of the thiolates.
\end{abstract}


\maketitle

 Nanoparticles consisting of metals  and oxides have been  shown to  exhibit magnetic behavior even    when    they are made only      of     non-magnetic     compounds \cite{Crespo2004,Rumpf2008,Crespo2006,Garcia2009,Garitaonanindia2008}.  When organic molecules are attached to Au via thiol bonds, Au  nanoparticles (NPs) with a size of 1.6 nm show a kind of ferromagnetic response, even at  room temperature.  The saturation  magnetization  is very  low,  being typically  in  the range  $10^{-1}$-$10^{-2}$  Bohr magnetons per atom, and is weakly dependent on temperature in the range from 4K up to room temperature.   The  low magnitude  of the  magnetic  moment is  not caused by the few magnetic  impurities present, because any increase in the number of impurities causes the magnetic signal to disappears \cite{Crespo2006}.  The results of both  X-ray magnetic circular  dichroism (XMCD) and Au M\"ossbauer spectroscopy  have shown that the Au atom possses a magnetic moment  \cite{Garcia2009,Garitaonanindia2008}, which was  previously thought to be due  to the spins of Au  d-holes introduced by the  ligands \cite{hernando2006}.   However,  the  observed  saturation  at  room temperature  and  the shape  of  the  magnetization  curve still  lack convincing explanation. A more detailed understanding of the nature of the gold-alkenethiol bond is therefore required. In particular, understanding of the interaction between alkenethiols and Au  NPs is currently one  of the  main challenges  in the study of their magnetic structure.

The contribution of the sp levels in Au NPs is  particularly significant, and it is  the filling of these levels that  explains the  effects of the electronic shells in  noble  metal  clusters in general and    in  gold   clusters  in particular \cite{alonso}.   Using a theoretical approach, it  has been shown that the  coverage of the gold clusters by thiols that have well-defined compositions is rather disordered,  and that the  charge transfer  depends strongly  on  the coordination of sulphur atoms \cite{hakkinen}.    A range of different coordination numbers must therefore be taken into account in any investigation. Alkenethiols with longer  chains on Au surfaces have been the focus of detailed theoretical studies \cite{Yourdshahyan} and the presence of several different metastable states in energy have been shown.   The  sulphur  atoms in the gold layers becomes arranged in a bridge  configuration for the ground state, and this must then  be  primarily considered in assesing their  magnetic  signal.  Other NPs such as ZnO capped with  a range of different organic molecules, e.g. amine, thiol and topo, also exhibit ferromagnetism at room temperature \cite{garcia2007}. The results of XMCD show that  the magnetic signal of ZnO  NPs is due to the  conduction band,  which is mainly formed from empty  4sp Zn  surface states \cite{chaboy2010}. The  d-electrons of  Zn are too  far away in  terms of energy and clearly make no contribution to the magnetic signal.  It has been shown theoretically that a  carbon oxide  CO molecule  on  ZnO surface  donates a single electron  to the  4sp  surface orbitals, which are initially empty \cite{Anderson1986}.  Although there  have  been  a number of  studies  of the  changes  in  the geometrical structure induced by alkenethiols in gold surfaces and clusters, to our knowledge no  studies have yet focused on the magnetism of the NPs covered  by such molecules.   The results of such  a  study would be  of  particular interest  when considered  toghether  the  effect  of  magnetism  on  the  electronic properties of gold nanoparticles covered by thiols.

An all-electron local-orbitals scheme, namely the ADF (Amsterdam Density Functional) method \cite{ADF}, was used herein for calculations involving density-functional-theory (DFT). The valence basis set was composed of the d gold, sulphur, carbon and  hydrogen states, which were expanded in triple zeta with two polarization functions at each atom. The relativistic effects of gold were modeled using the Zeroth Order Regular Approximation (ZORA) and its scalar relativistic version was used for the structural optimizations. The data presented herein were obtained using generalized gradient density approximations (GGA) with the parametrized exchange-correlation functional of Perdew-Burke-Ernzerhof according to Ref. \cite{perdew}. The optimization of the geometry was carried out until the forces were all smaller than 0.02 eV/\AA.   The magnetic anisotropy (MAE) was obtained by means of self-consistent ZORA relativistic calculations, including spin-orbit. The resulting GGA data were regarded as a lower estimate for the MAE. In order to obtain an upper estimate, an orbital polarization correction (OPC) \cite{eriksson} should be considered, but this work is beyond the scope of the present study.

The singly or doubly occupied levels of the Au clusters are responsible for the odd-even oscillations. These oscillations do not vary strongly much with the number of atoms in the cluster. We have herein therefore chosen to investigate the interaction between alkenethiols $-SCH_3$ and minimal gold clusters. The interactions of gold clusters with thiols are considered to be a suitable model of adsorption of thiols onto larger nanoparticles \cite{hakkinen}.

\begin{figure}[h]
\includegraphics[width=5.300in]{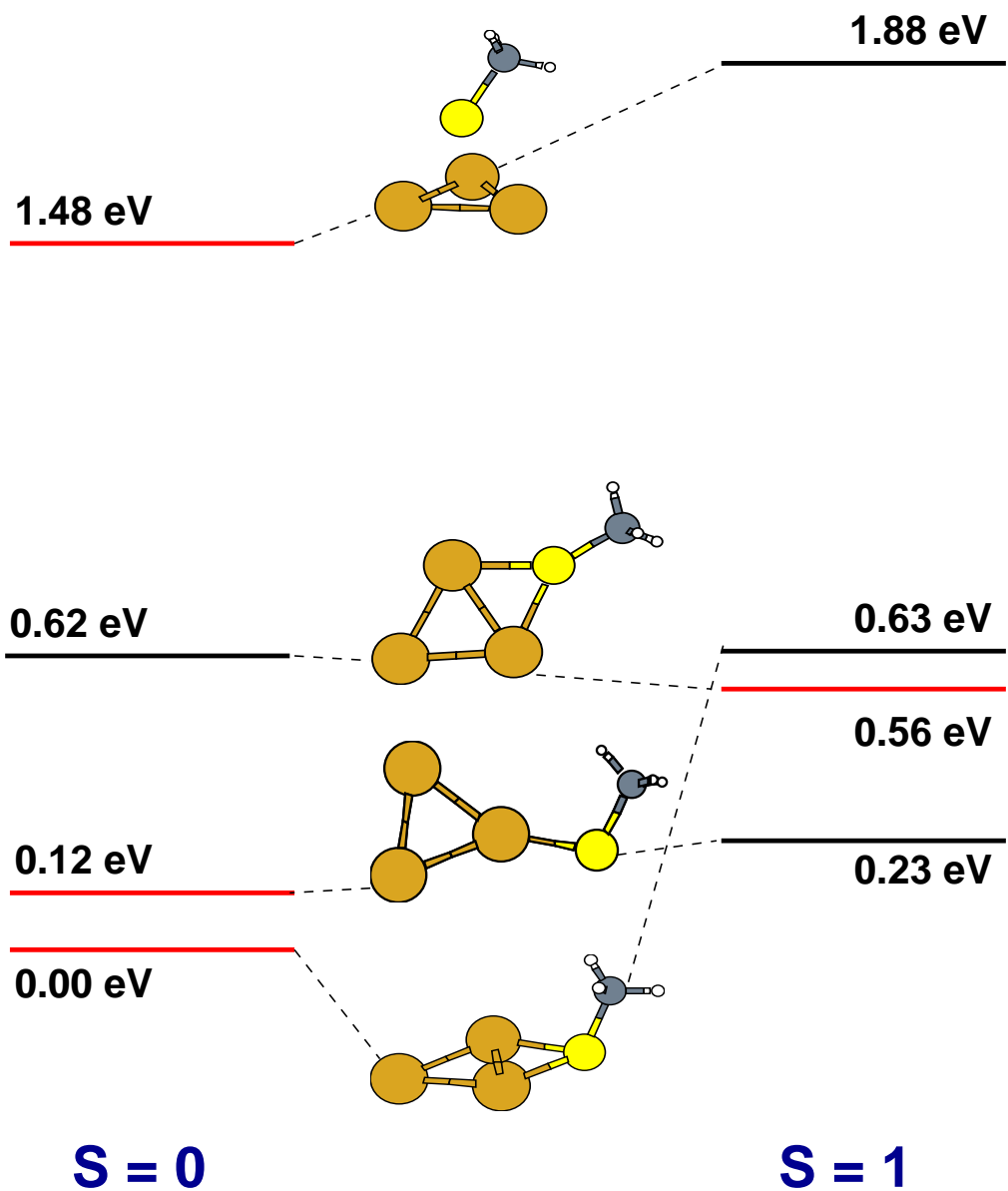}
\caption{ DFT energies for various states of spin magnetic moments and thiol-$Au_3$ configurations. The energies refer to the spin and geometry of the ground state. In order of size, the atoms are gold, sulphur, carbon, and hydrogen. Note the different types of Au-sulfur bonds.
}
\label{fig1}
\end{figure}

We begin our analysis with a cluster of $Au_3$, which allows us to investigate a number of different coordination numbers for the sulphur atom. $Au_3SCH_3$ has an even number of electrons, and thus has a spin moment equivalent to an even number of Bohr magnetons. In order to determine the lowest-energy geometry and the spin magnetic state, four possible high-symmetry structures were optimized for each of the possible values of low spin $\mu_S=$0, 1, 2 etc. We  only compared the data for $\mu_S=$0 and 1 in Fig. \ref{fig1}, because the other spin states produce an even higher total energy. The point group symmetries  were fixed in our calculations to reproduce those of the geometries in the surfaces of the nanoparticles. The geometry of the gold atoms was not fixed because the adsorbates of thiols on gold can alter the atomic distances of Au-sulphur and Au-Au binding.

 Calculations using DFT yield energies that depend on the exchange-correlation approach used. In order to confirm the qualitative validity of the calculated energies, structural sequence, and spin states of Fig.  \ref{fig1}, we performed calculations with many other functionals, as implemented in ADF, using the fixed geometry obtained with the previous GGA approach. We obtain that the energetic difference between the different spin states was sometimes higher [though never lower] than that obtained from the previous GGA values using the PBE approach. The other functionals produced energetic differences of about 1.4 eV between the non- and spin-polarized structure. These supporting calculations therefore confirm the main findings obtained using PBE, in that the bonds of an $SCH_3$ molecule with a three-atom cluster form the structure depicted in the bottom of Fig. \ref{fig1} which has a total spin of zero.

Our findings show a gain in total energy of 0.12 eV for the $-SCH_3$ adsorbate with the gold cluster in a bridge position, this being the ground-state structure with respect to the molecule in the $Au_3$ plane. We are not aware of any previous investigation of this structural form of $Au_3SCH_3$. Although other molecules of Au-thiols have been studied previously \cite{parrinello}, no report has been made of a model of the surface of gold nanoparticles. Evidence for such 'tilt' arrangement for the molecule was previously provided by Ref. \cite{Yourdshahyan} for thiols on gold surfaces. Furthermore, thiols have theoretically been predicted \cite{Yourdshahyan} to form a mat on a gold surface with perpendicularly oriented molecules. This is given by the bridge position with all the atoms bonded to S in the same plane, as shown in the third geometry of Fig. \ref{fig1}. However, this form has a larger energy than the tilt case, because it is a metastable position by 0.56 eV. Althought this form has a similar coordination number to that of the studied case of $C_6H_5-S-Au_{13}$ \cite{mujica}, is not considered in the results that follow, despite its spin polarization for alkanethiols.

\begin{figure}[h]
\includegraphics[width=5.300in]{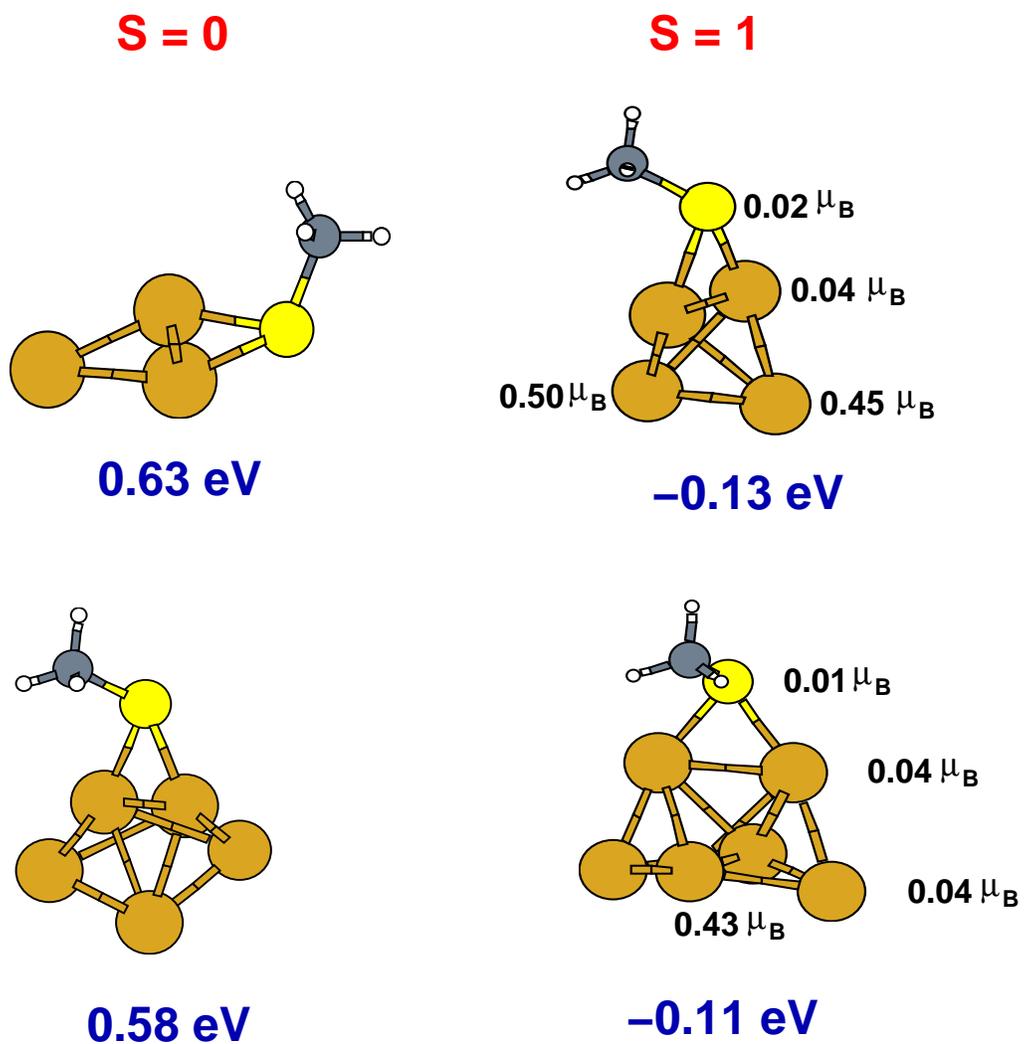}
\caption{ Relaxed structures for sulphur-Au$_n$ clusters. The numbers below the structures refer to the energetic difference between spin polarized and spin compensated structure. The numbers on the atoms refer to the local spin densities on the atoms for the ground states with spin polarization. Note that the odd-even oscillations in spin are shifted by 1 compared to the bare gold clusters.  
}
\label{fig2}
\end{figure}

The structural forms of other gold cluster structures are given in Fig. \ref{fig2}. The panel on the far right hand side shows the structures of S=0, as previously discussed for the ground state of $Au_3$-thiol, while the left-hand panel refers to clusters with spin polarization (S=1). The numbers shown below the structures give the spin polarization energy in each case, which are almost independent of the size of the cluster. The related energy differences $\Delta E$ are around 0.5 eV and -0.2 eV for the odd and even cases, respectively.  These values should be considered to be lower estimates for the polarization, because they were evaluated without the so-called orbital polarization. The most important characteristic is the shift between odd-even oscillations, by a single gold atom, with respect to bare gold clusters. The structure is spin compensated for an even number of gold atoms in the thiol-cluster, while it is spin polarized otherwise. 

It is perhaps somewhat surprisingly that the bonding of thiols on gold clusters does not lead to the elimination of the spin magnetic moment, but rather to a new paradigm, i.e. the spin moments induced by the thiols are unchanged, as in free particles. It is noteworthy that our spin polarized state can be more stable in energy, by $>$0.2 eV, than the state withouth spin polarization. In consequence, the level of the highest occupied molecular orbital HOMO is singly occupied, and follows the local magnetization on atoms. Further analysis of this characteristic for $Au_nSCH_3$ clusters that contain an even number of gold atoms, as also seen in Fig. \ref{fig2}, show that, while the local magnetic moments of the thiols nearest the gold atoms are mostly negligible ($<$ 0.05 $\mu_B$ per atom), those of the lower gold atoms ($\sim$0.5 $\mu_B$ per atom) are even higher than the experimental values. These very high values of  magnetic moment per atom may also be accompanied by large ground-state orbital moments, of about 1 $\mu_B$ per molecule. The spin moments are largely suppressed when the thiols are located parallel to the gold surface. Depending on the orientation of the molecule, the spin moment varies due to different numbers of thiol-gold bonds. We have herein neglected variations in the bond distances.

\begin{figure}[h]
\includegraphics[width=4.000in]{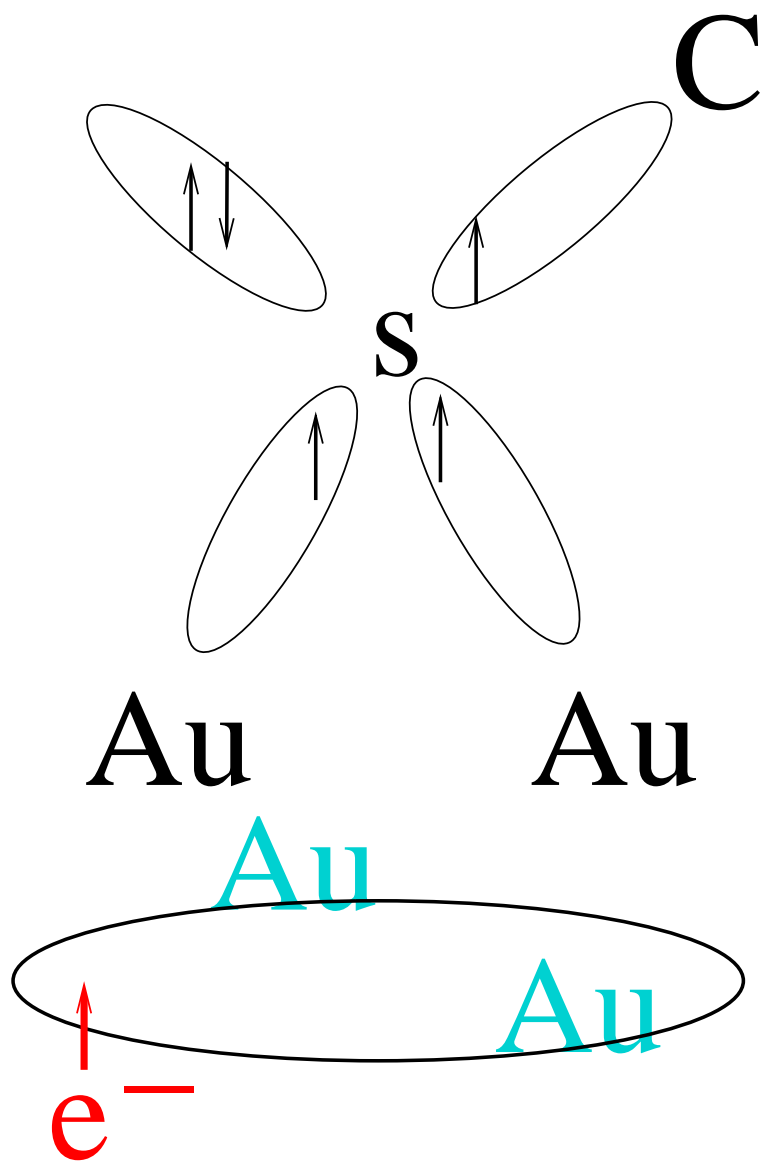}
\caption{ Bonding mechanism of thiols on gold metal. The orbitals around the sulphur denote $sp^3$ like configurations and the arrows indicate electrons from a sulphur atom. In order to form bonds with two gold Au atoms on a surface, a single sp electron must  pass from sulphur to the metal cluster.
}
\label{fig3}
\end{figure}

Although some of the gold-thiol molecules described herein were investigated both experimentally \cite{Crespo2004}, by quantum mechanical calculations \cite{parrinello}, and using spin-polarized GGA \cite{hakkinen}, their behaviour in terms of their magnetic moment merits further study.  Figure \ref{fig3} shows the bonding mechanism of thiolates on metal clusters. A sulphur atom is shown here joined to two Au atoms with $sp^3$ hybridization. The spin population is smaller in these sulphur and bonded gold atoms than in that of the next nearest neighbors below the sulphur-gold bonds. There are five sulphur electrons in total: the two bonds with gold atoms share two electrons, an electron is shared in the bond with the carbon atom, and two more form the lone pair in the $sp^3$ hybridization. Because each sulphur atom contributes with a total of six electrons, a single electron is left over and is passed to the gold cluster. It is this sp electron that contributes to the spin polarization.

A related discussion that supports the back-donation of a sp electron from the thiols to gold arises from a comparison between the eletronegativities of thiols and those of their gold counterparts. We calculated the ionization potentials (IPs) and eletroaffinities (EAs), thereby obtaining the Mulliken electronegativities  $M=(IP+EA)/2$. This M value is the the negative of the electrochemical potential. The electronegativity of the $SCH_3$ unit is slightly higher (5.43 eV) \cite{note} than the work function of around 5.22 eV for the Au surface \cite{yu} and is comparable with the M value for gold nanostructures ($\lessapprox$~6~eV) \cite{ayuela}. The $SCH_3$ unit behaves rather differently from the sulphur atom (M=6.21 eV). In consecuence, the transfer of charge from gold to sulfur cannot be assumed a priori. In fact, we stress that independently of the partition scheme used, the calculated transfer of charge between the thiols and the gold clusters is less than 0.1 $e$. Such a charge neutrality is also seen in other systems, even in those assumed to be highly ionic \cite{raebiger}, and supports the previously commented mechanism of back-donation.  

We now compare our findings with those obtained from experiments on thiols and gold nanoparticles. To this end, it must be stressed that the sulphur atom is not negatively charged, even though XANES experiments have shown that the 5d orbitals of gold are slightly open. The sp contribution to the magnetic signal is larger than that of the d orbitals. The back-donated electron from sulphur to gold also sits in the sp states of the Au bond, and cannot be differentiated by the previous results using XANES method. These sp electrons are almost free in the second gold layer and could orbit, and it is these that are responsible for the magnetism. These findings are consistent with experimental results in which a number of gold layers are needed in order to engender thiol induced magnetism. This general picture is in agreement with the donor mechanism of carbon monoxide to Zn atoms on a ZnO surface \cite{Anderson1986}, in that there is donation of a single electron from the molecule to the 4sp surface orbitals of Zn, which are initially empty. More importantly, it is also in agreement with the 4sp magnetism observed by XMCD in thiol capped ZnO nanoparticles \cite{chaboy2010}. 

The method of application of thiols is via their deposition in layers on the surface of gold nanoparticles. The magnetism that results should be that of the electrons donated by  the capping molecules, which are confined to a thin layer below the surface, and form a two-dimensional electron gas. 
We have herein determined the means by which this surface layer is unfilled, thereby giving rise to a permanent magnetic moment \cite{hernando}. Orbital and spin angular moments are formed at the generally unfilled Fermi level of the system, which is confined to a spherical shell at the surface. The unfilled Fermi level of the surface band is characterized by a collective magnetic moment that is rather similar to an atomic orbital, but with a much larger quantum number. The order of magnitude (0.1 or 0.01 $\mu_B$ per atom) is in agreement with that measured experimentally  \cite{Crespo2004,Rumpf2008,Crespo2006,Garcia2009,Garitaonanindia2008}.


Among all the possible combinations of thiol and noble metals,  we have herein described our investigation of  Au clusters, but it should be mentioned that copper shows similarly good bonding behavior with thiols. The related magnetic ground states have the same spin state (1 $\mu_B$) and lie at similar energies with respect to the state with zero  moment. This finding is similar to recent experimental work carried out for these nanoparticles \cite{Garitaonanindia2008} and deserves further investigation.

By including spin-orbit coupling, we find that thiols attached ot $Au_4$ and $Au_6$ show a magnetic anisotropies  MAE of 1.43 and 0.98 meV per $SCH_3$ molecule, respectively, while the change of MAE in the plane parallel to the surface is less than an order of magnitude smaller. These figures are comparable with the giant magnetic anisotropy of single cobalt atoms on a Pt surface (9 meV per Co atom) \cite{gambardella}. Such large values of MAE demonstrate the stability of the magnetism discussed herein at room temperature, and  are a factor of ten times larger than the equivalent values for tetragonal Ni (0.2 meV). The results of a recent experiment \cite{Crespo2004} showed that the magnetism of gold nanoparticles covered by thiols can indeed survive almost unchanged at room temperature, and it could therefore be used in hyperthermic applications.
 
In summary, we predict that the bonding of alkanethiols on gold clusters (i.e.$SCH_3Au_n$) results in a bridge arrangement with a deviated axis in a polarised ground  state for even n values of n. This structure is characterized by a division of functions, in that while the sulphur atom and the close pair of atoms are responsible for the chemical  bonding, it is the outer gold atoms that host the larger part of the magnetic moment. The large magnetic moment is preserved in the structure of the nanoparticles because they have lower symmetries. Thus, it should be possible for the particles of other noble metals to show similar magnetic behavior. Further avenues of investigation could include the fabrication of separated thiols on the surface of nanoparticles, and the development of novel magnetic nanoparticles for medical applications. 

{\it Acknowledgement.} We gratefully acknowledge the support of the Basque Departamento de Educaci\'on and the UPV/EHU (Grant No. IT-366-07), the Spanish Ministerio de Innovaci\'on, Ciencia y Tecnolog\'{\i}a (Grant  No. MAT2009-14741-C02-01, CSD2007-00010, and FIS2007-66711-C02-02),  and  the ETORTEK  research program funded  by  the  Basque Departamento de Industria and the Diputaci\'on Foral de Guip\'uzcoa.

\bibliography{draft_aps.bib}

\begin{thebibliography}{23}
\bibitem [{not()}]{note}%
  \BibitemOpen
  \href@noop {} {\bibinfo  {journal} {Longer chains vary their
  electronegativity in the order of tens of eV, therefore our conclusion fully
  hold.}\ }\BibitemShut {NoStop}%
\end{thebibliography}

\end{document}